\documentclass[%
 showpacs,
 showkeys,
 amsmath,amssymb,
]{revtex4-1}

\usepackage{graphicx}
\usepackage{dcolumn}
\usepackage{bm}

\usepackage{CJK}

\usepackage{color}
\usepackage{amssymb}
\usepackage{amsfonts}

\usepackage[colorlinks,urlcolor=blue,linkcolor=blue,citecolor=blue,anchorcolor=blue]{hyperref}

\begin{document}


\preprint{APS/123-QED}

\title{Coherent and incoherent nonparaxial self-accelerating Weber beams}

\author{Yiqi Zhang$^1$}
\email{zhangyiqi@mail.xjtu.edu.cn}
\author{Junfeng Liu$^1$}
\author{Feng Wen$^1$}
\author{Changbiao Li$^1$}
\author{Zhaoyang Zhang$^1$}
\author{Yanpeng Zhang$^{1}$}
\author{Milivoj R. Beli\'c$^{2}$}
\affiliation{%
 $^1$Key Laboratory for Physical Electronics and Devices of the Ministry of Education \& Shaanxi Key Lab of Information Photonic Technique,
Xi'an Jiaotong University, Xi'an 710049, China \\
$^2$Science Program, Texas A\&M University at Qatar, P.O. Box 23874 Doha, Qatar 
}%

\date{\today}

\begin{abstract}
  \noindent
  We investigate the coherent and incoherent nonparaxial Weber beams, theoretically and numerically.
  We show that the superposition of coherent self-accelerating Weber beams with transverse displacement cannot display the nonparaxial accelerating Talbot effect.
  The reason is that their lobes do not accelerate in unison,
  which is a requirement for the appearance of the effect.
  While for the incoherent Weber beams, they naturally cannot display the accelerating Talbot effect but can display the nonparaxial accelerating properties,
  although the transverse coherence length is smaller than the beam width,
  based on the second-order coherence theory.
  Our research method directly applies to the nonparaxial Mathieu beams as well,
  and one will obtain similar conclusions as for the Weber beams,  although this is not discussed in the paper.
  Our investigation identifies families of nonparaxial accelerating beams that do not exhibit the accelerating Talbot effect,
  and in addition broadens the understanding of coherence properties of such nonparaxial accelerating beams.
\end{abstract}

\maketitle

%
\section{Introduction}

The paraxial wave propagation equation is formally equivalent to the Schr\"odinger equation in quantum mechanics \cite{berry.ajp.47.264.1979},
which helped a lot in the discovery of nondiffracting self-accelerating Airy beams \cite{siviloglou.ol.32.979.2007,siviloglou.prl.99.213901.2007}.
In the past decade, investigations on the topic were many and acquired a rapid development.
Until now, studies of Airy beams have been reported in nonlinear media \cite{kaminer.prl.106.213903.2011,dolev.prl.108.113903.2012,zhang.ol.38.4585.2013,zhang.oe.22.7160.2014},
optical fibers \cite{zhang.oe.23.2566.2015,hu.prl.114.073901.2015}, and other systems.
To manipulate the behavior of Airy beams, different external potentials such as harmonic potential \cite{bandres.oe.15.16719.2007,zhang.oe.23.10467.2015,zhang.aop.363.305.2015}
or a linear potenial \cite{liu.ol.36.1164.2011,efremidis.ol.36.3006.2011}, have been explored.

To recall, the paraxial wave propagation results from the Maxwell's equations after applying the slowly-varying envelope approximation.
Without such an approximation, one investigates nonparaxial optical problems using the Helmholtz equation
that follows directly from the Maxwell's equations \cite{gutierrez-vega.josaa.22.289.2005}.
It has been demonstrated that the solutions of two-dimensional (2D) Helmholtz equation are plane waves in Cartesian coordinates,
Bessel beams in circular cylindrical coordinates \cite{alonso.ol.37.5175.2012,kaminer.prl.108.163901.2012,lumer.optica.2.886.2015,zhang.arxiv1603.08339.2016},
Mathieu beams in elliptic cylindrical coordinates \cite{zhang.prl.109.193901.2012,aleahmad.prl.109.203902.2012},
and Weber beams in parabolic cylindrical coordinates \cite{bandres.ol.29.44.2004,lopez-mariscal.oe.13.2364.2005,kartashov.ol.33.141.2008,zhang.prl.109.193901.2012,bandres.njp.15.013054.2013}.
In addition to the 2D case, one can also solve the 3D Helmholtz equation
for the 3D accelerating beams utilizing the Whittaker integral \cite{alonso.ol.37.5175.2012,bandres.oe.21.13917.2013,zhang.epl.107.34001.2014}.
The reason why investigations of accelerating beams are so hot at the moment is their numerous potential applications,
such as in particle manipulation \cite{baumgartl.np.2.675.2008,zhang.ol.36.2883.2011,schley.nc.5.5189.2014},
electron beam shaping \cite{voloch.nature.494.331.2013},
and super-resolution imaging \cite{jia.np.8.302.2014}, to name a few.
Recent studies indicated that accelerating beams have opened a new window in explorations of elusive problems in
general relativity \cite{bekenstein.np.11.872.2015,sheng.nc.7.10747.2016} and quantum particles \cite{kaminer.np.11.261.2015}.
For more details on accelerating beams, the reader may consult review articles \cite{hu.book.2012,bandres.opn.24.30.2013,levy.po.2016} and references therein.

In this article, we investigate the coherent and incoherent nonparaxial accelerating Weber beams.
We focus on the two aspects:
(1) Can coherent Weber beams exhibit nonparaxial accelerating Talbot effect, as Bessel beams do?
(2) What behavior will incoherent Weber beams display? Will they lose the nonparaxial accelerating property?
The organization of the article is as follows.
In Sec. \ref{model}, we introduce the theoretical model and the Weber beams in parabolic cylindrical coordinates.
In Sec. \ref{coherent}, we investigate the coherent Weber beams and check whether the nonparaxial accelerating Talbot effect can take place or not.
In Sec. \ref{incoherent}, incoherent Weber beams and their characterization is considered, including the degree of coherence and the coherence length.
We conclude the article in Sec. \ref{conclusion}.

\section{Theoretical model}\label{model}

We begin from the two-dimensional Helmholtz equation, which can be written as
\begin{equation}\label{eq1}
  \left( \frac{\partial^2}{\partial x^2} + \frac{\partial^2}{\partial z^2} \right) \psi + k^2 \psi = 0,
\end{equation}
where $k$ is the wavenumber.
One of the shape-preserving solutions of Eq. (\ref{eq1}) is the so-called Weber wave function in the parabolic cylindrical coordinates.
The transformation relation between the Cartesian coordinates $(x,z)$ and the parabolic cylindrical coordinates $(\eta,\xi)$ is
$x+iz=(\eta+i\xi)^2/2$ with $\eta\in(-\infty,\infty)$ and $\xi\in[0,\infty)$.
By utilizing variable separation, that is, by writing the solution of Eq. (\ref{eq1}) as $\psi(\xi,\eta)=R(\xi)\Phi(\eta)$, one obtains
two ordinary differential equations \cite{zhang.prl.109.193901.2012,bandres.njp.15.013054.2013},
\begin{subequations}\label{eq2}
\begin{align}\label{eq2a}
\frac{\partial^2 R(\xi)}{\partial \xi^2} + \left( k^2\xi^2-2ka \right) R(\xi)=0,
\end{align}
\begin{align}\label{eq2b}
\frac{\partial^2 \Phi(\eta)}{\partial \eta^2} + \left( k^2\eta^2+2ka \right) \Phi(\eta)=0,
\end{align}
\end{subequations}
where $2ka$ is the separation constant.
If one sets $\sqrt{2k}\xi=v$ and $\sqrt{2k}\eta=u$,
Eqs. (\ref{eq2a}) and (\ref{eq2b}) transform into the canonical form of the parabolic cylinder differential equations
\begin{subequations}\label{eq3}
\begin{align}\label{eq3a}
\frac{\partial^2 R(v)}{\partial u^2} + \left( \frac{v^2}{4} - a \right) R(v)= &0,
\end{align}
\begin{align}\label{eq3b}
\frac{\partial^2 \Phi(u)}{\partial v^2} + \left( \frac{u^2}{4} + a \right) \Phi(u)= &0.
\end{align}
\end{subequations}

The solutions of Eqs. (\ref{eq2a}) and (\ref{eq2b}) are determined by the same Weber functions,
but the corresponding eigenvalues have opposite signs.
If we denote the even and odd solutions of Eq. (\ref{eq2a}) as $P_e$ and $P_o$,
the final even and odd transverse stationary solutions of Eq. (\ref{eq1}) in parabolic cylindrical coordinates are expressed as
\begin{subequations}\label{eq4}
\begin{align}\label{eq4a}
W_e(x,z;a)=\frac{1}{\sqrt{2}\pi}|\Gamma_1|^2 P_e\left(\sqrt{2k}\xi;a\right) P_e\left(\sqrt{2k}\eta;-a\right),
\end{align}
\begin{align}\label{eq4b}
W_o(x,z;a)=\frac{2}{\sqrt{2}\pi}|\Gamma_3|^2 P_o\left(\sqrt{2k}\xi;a\right) P_o\left(\sqrt{2k}\eta;-a\right),
\end{align}
\end{subequations}
respectively, where
\begin{align}\label{eq5}
  P_{e,o}(t,a)= \sum^\infty_{n=0} c_n \frac{t^n}{n!},
\end{align}
$\Gamma_1 = \Gamma[(1/4)+(1/2)ia]$, $\Gamma_3=\Gamma[(3/4)+(1/2)ia]$,
and the coefficients $c_n$ satisfy the following recurrence relation:
\begin{align}\label{eq6}
  c_{n+2} =  ac_n - \frac{n(n-1)c_{n-2}}{4}.
\end{align}
For $P_e$ ($P_o$), the first two $c_n$ coefficients are $c_0=1$ and $ c_1=0$ ($c_0=0$ and $ c_1=1$) \cite{bandres.ol.29.44.2004}.
Therefore, the general transverse stationary solution of Eq. (\ref{eq1}) can be written as
\begin{equation}\label{eq7}
  W(x,z;a)=W_e(x,z;a)+iW_o(x,z;a).
\end{equation}

\begin{figure}[htbp]
\centering
  \includegraphics[width=0.75\columnwidth]{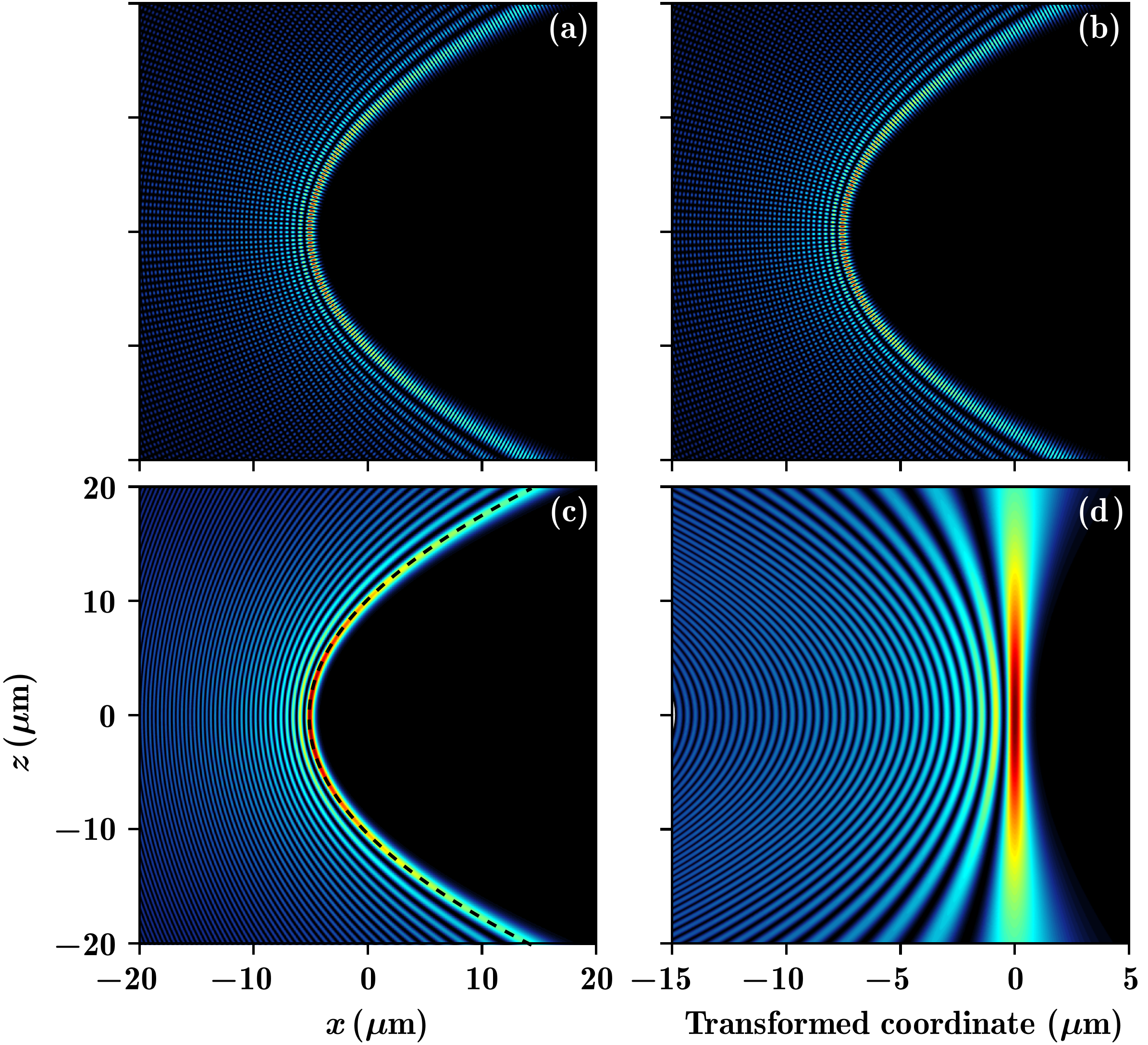}
  \caption{Intensity distributions of the even (a), odd (b), and general (c) transverse stationary solutions.
  The dashed curve in (c) is the theoretical trajectory of the main lobe.
  (d) Same as (c), but the transverse coordinate is transformed according to Eq. (\ref{eq8}).
  (a)-(c) share the same scales, and all the panels share the same vertical scale.
  Other parameters are $a=50$ and $\lambda=600$ nm.
  }
  \label{fig1}
\end{figure}

According to \cite{bandres.njp.15.013054.2013}, if $a\ge10$,
the trajectory of the main lobe of the Weber beam in Eq. (\ref{eq7}) is of the form
\begin{equation}\label{eq8}
  x = \frac{1}{2f(a)}z^2 - \frac{1}{2}f(a),
\end{equation}
where \[f(a) \approx \frac{1}{2k} \left(2\sqrt{a} + \frac{0.9838}{\sqrt[25]{a^4}} \right)^2.\]
Clearly, the trajectory in Eq. (\ref{eq8}) is a parabolic curve that depends on the value of $a$.
Therefore, if one fixes the value of $a$, then the trajectory is also determined regardless of whether there is a transverse displacement or not.
To elucidate the Weber beams clearly, we display in Fig. \ref{fig1} the intensity distributions of the even, odd and general transverse stationary solutions.
In Fig. \ref{fig1}(c), the theoretical trajectory of the main lobe is also exhibited, as indicated by the dashed curve.
However, it should be stressed that lobes of the Weber beam do not accelerate in unison.
That is, the path differences along different lobe trajectories vary such that they are not uniform and fixed.
This is unlike the nonparaxial Bessel beams, where the lobes follow circular trajectories.
The rate of change of the path length with the radial distance there is constant.
A more clear sketch of the Weber beam is displayed in Fig. \ref{fig1}(d),
in which the transverse coordinate is transformed according to Eq. (\ref{eq8}) that describes the trajectory of the main lobe.
One sees that all lobes except the main lobe do not propagate along a straight trajectory.

In the following, we discuss the coherent and incoherent Weber beams in more detail, based on the basic theory introduced here.

\section{Coherent Weber beams}\label{coherent}

Due to the linear property of Eq. (\ref{eq1}), the solution in Eq. (\ref{eq7}) with an arbitrary transverse displacement $\Delta x$ is also a solution.
Thus, the general solution of Eq. (\ref{eq1}) in parabolic cylindrical coordinates can be written as
\begin{equation}\label{eq9}
  W(x,z;a) = \sum_{a\in\mathbb{R}} \sum_{n\in\mathbb{Z}} \left[ W_e(x+n\Delta x,z;a)+iW_o(x+n\Delta x,z;a) \right],
\end{equation}
in which the coefficients of all the components are set to be 1.

\begin{figure}[htbp]
\centering
  \includegraphics[width=0.75\columnwidth]{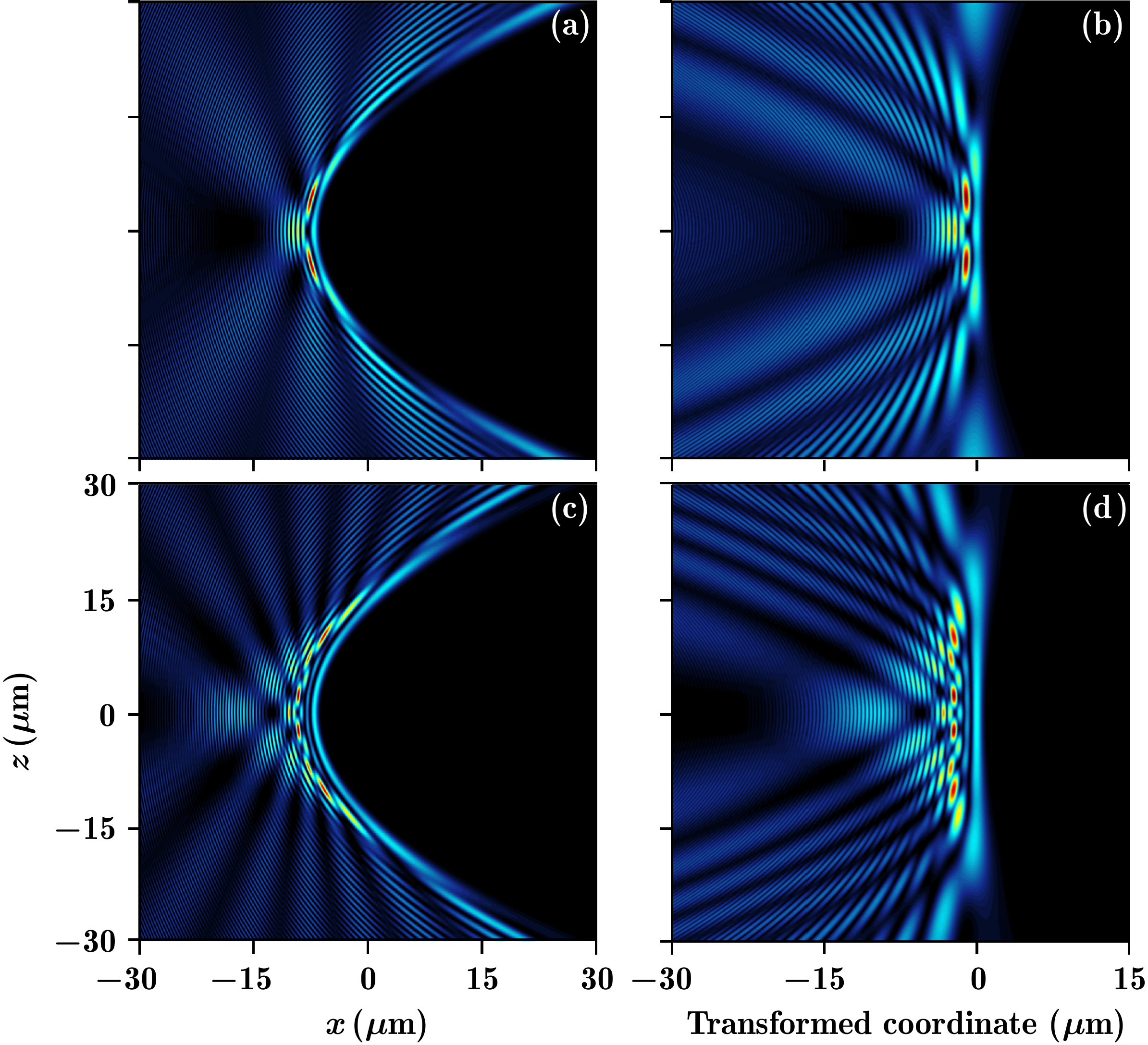}
  \caption{(a) Intensity distributions of the general nonparaxial Weber beam with two components with $n = 0$ and $n=3$, respectively.
  (b) Same as (a) but the transverse coordinate is transformed.
  (c) and (d) Setup is as (a) and (b), but for $n = 0$ and $n=6$, respectively.
  (a) and (c) share the same scales, while (b) and (d) share the same scale.
  Other parameters are $a=70$, $\lambda=600$ nm, and  $\Delta x \approx 39.07~{\rm nm} $.
  }
  \label{fig2}
\end{figure}

Even though the nonparaxial Weber beam can be reduced to the paraxial Airy beam under proper conditions \cite{zhang.prl.109.193901.2012,bandres.njp.15.013054.2013},
the lobes of the Weber beam do not accelerate in unison,
as is visible in Fig. \ref{fig1}(c) and from the ``periodic'' Weber beam in Fig. 4(b) of \cite{bandres.njp.15.013054.2013}.
We would like to explain here in details.
The parabolic coordinate system is defined by confocal parabolae
and each parabola with different latus rectum and the separation between lobes varies along the parabolae.
On the other side, the lobes in Airy beams are governed by one single parabola.
A single Airy beam as it propagates maintains its profile within the region of existence,
but it is not the case for Weber beams since their lobes change their separation as they propagate.

Since the theoretical trajectory of the main lobe depends on $a$, the Weber beams with different $a$ do not accelerate in unison.
Therefore, we only consider the solution with a fixed $a$ and arbitrary transverse displacement.
Then, Eq. (\ref{eq9}) is reduced into
\begin{equation}\label{eq10}
  W(x,z;a) = \sum_{n\in\mathbb{Z}} \left[ W_e(x+n\Delta x,z;a)+iW_o(x+n\Delta x,z;a) \right].
\end{equation}

We display the intensity distribution of the beam with only two components given by Eq. (\ref{eq10}) in Fig. \ref{fig2}(a)
and in Fig. \ref{fig2}(b) with a transformed coordinate, according to Eq. (\ref{eq8}).
One finds that the beam exhibits a quasi-periodic behavior along the $z$ direction,
however this pattern cannot be called a Talbot effect \cite{lumer.prl.115.013901.2015,zhang.ol.40.5742.2015}.
The reason is clearly -- the lobes do not accelerate in unison,
and the unison acceleration is a requirement \cite{kaminer.prl.108.163901.2012} for the Talbot effect.
What one sees is just the interference pattern of limited periodicity.
Another reason is that the intensity and the width of each lobe respectively decreases  and increases with the increasing propagation distance $z$,
as elucidated in Fig. \ref{fig1}(d).
Clearly, the nondiffracting properties of the nonparaxial Weber beams are much worse than those of the nonparaxial Bessel beams.
In the Bessel beams one can see an extended region in which the nonparaxial Talbot effect is observed.
Due to the above reasons, we assert that one cannot obtain nonparaxial accelerating Talbot effect based on the Weber beams.
As a result, different from the nonparaxial Bessel beams \cite{zhang.arxiv1603.08339.2016} and Airy beams \cite{zhang.ol.40.5742.2015},
the nonparaxial Weber beams do not possess duality.

Even though one cannot obtain the nonparaxial Talbot effect, there is indeed quasi-periodicity in the general transverse stationary solution
caused by interference,
and the pattern is also dependent on the transverse displacement $\Delta x$.
For comparison, we display the quasi-periodic Weber beam in Figs. \ref{fig2}(c) and \ref{fig2}(d) again
that correspond to Figs. \ref{fig2}(a) and \ref{fig2}(b), but with a bigger $\Delta x$.
One finds that the interference fringes are denser in Figs. \ref{fig2}(c) and \ref{fig2}(d) than those in Figs. \ref{fig2}(a) and \ref{fig2}(b),
which indicates the quasi-periodic pattern has a smaller ``period'' if $\Delta x$ is larger.

\section{Incoherent Weber beams}\label{incoherent}

For the spatially incoherent case, the intensity of the beam which is composed of an arbitrary number of independent modes, should be written as
\begin{equation}\label{eq11}
  I(x,z;a) = \sum_{a\in\mathbb{R}} \sum_{n\in\mathbb{Z}} \left| W_e(x+n\Delta x,z;a)+iW_o(x+n\Delta x,z;a) \right|^2.
\end{equation}

\begin{figure}[htbp]
\centering
  \includegraphics[width=0.8\columnwidth]{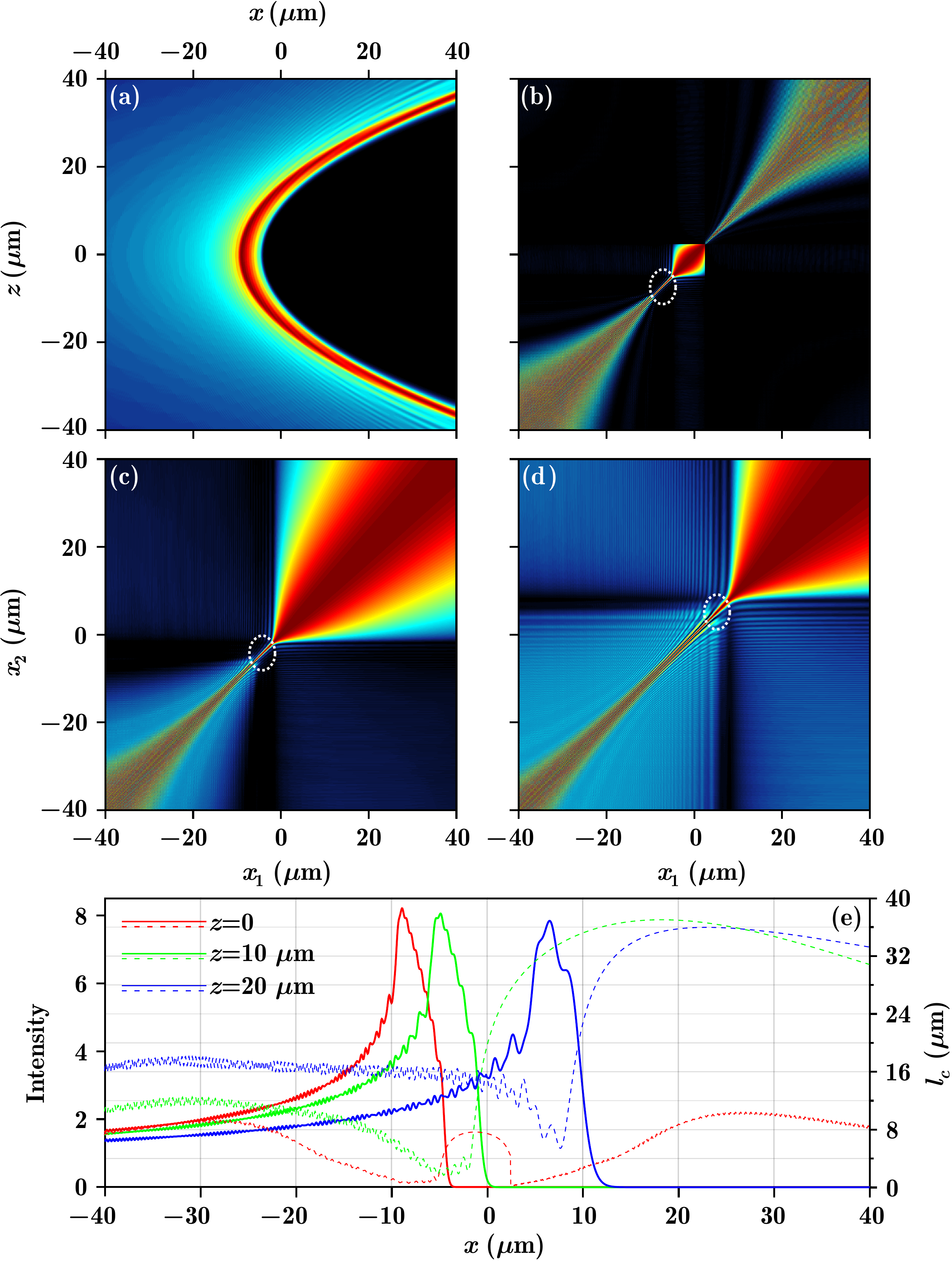}
  \caption{(a) Intensity distribution of an incoherent Weber beam composed of 129 modes with random transverse displacements in the range $[-2.5~\mu{\rm m},~2.5~\mu{\rm m}]$.
  (b)-(d) Complex degree of spatial coherence function $|\gamma_{12}(x_1,x_2)|^2$ at $z=0$, $z=10~\mu$m, and $z=20~\mu$m, respectively.
  They share the same variables and scales.
  The dotted ellipses indicate the effective regions of the coherence function.
  (e) Intensity distributions and transverse coherence lengths of the incoherent Weber beam corresponding to (b)-(d).
  Solid and dashed curves refer to the left and right vertical coordinates and variables, respectively.
  Other parameters are $a=70$ and $\lambda=600$ nm.
  }
  \label{fig3}
\end{figure}

One can check the coherence characteristics of the beam represented in Eq. (\ref{eq11})
through the complex degree of coherence \cite{wolf.book.2007}
\begin{equation}\label{eq12}
  \gamma_{12} \left( \vec{r}_1,\vec{r}_2 \right) = \frac{\Gamma \left( \vec{r}_1,\vec{r}_2 \right)}{\sqrt{I \left( \vec{r}_1 \right)} \sqrt{I \left( \vec{r}_2 \right)}},
\end{equation}
where $\Gamma \left( \vec{r}_1,\vec{r}_2 \right)$ is the mutual coherence function, defined as
\begin{equation}\label{eq13}
  \Gamma \left( \vec{r}_1,\vec{r}_2 \right) = \sum_{n\in\mathbb{Z}} W_n^* \left( \vec{r}_1 \right) W_n \left( \vec{r}_2 \right),
\end{equation}
where $W_n$ represent the beam components and asterisk denotes the complex conjugate.
It should be recalled that
\begin{equation}\label{eq14}
0 \le \left| \gamma_{12} \left( \vec{r}_1,\vec{r}_2 \right) \right| \le 1,
\end{equation}
where 0 represents a completely incoherent and 1 a completely coherent beam.
Connected with the coherence function, it is also convenient to explore the coherence properties via
the transverse coherence length, which is defined as
\begin{equation}\label{eq15}
  l_c \left( x \right) = \int_{-\infty}^{+\infty} \left| \gamma_{12} \left( x, x' \right) \right|^2 d x'.
\end{equation}

Based on Eq. (\ref{eq11}), the intensity of the incoherent Weber beam is exhibited in Fig. \ref{fig3}(a).
Since all the components accelerate along the same trajectory (a parabolic curve with transverse displacements),
one can still observe the nonparaxial acceleration property of the incoherent beam,
even though the oscillating lobes disappear due to the superposition of incoherent modes.
Intensity distributions of the beam at certain distances are displayed as the solid curves in Fig. \ref{fig3}(e),
from which one finds that the beam indeed accelerates nonparaxially with a nearly preserved shape.
Note that the intensity of the beam depends on the number of modes, and the larger the number the higher the intensity.

To investigate the incoherence of the beam, in Figs. \ref{fig3}(b)-\ref{fig3}(d),
we present the corresponding complex degree of spatial coherence function,
according to Eq. (\ref{eq13}), at $z=0$, $z=10~\mu \rm m$, and $z=20~\mu \rm m$, respectively.
The dashed ellipses show the spatial location of the main lobe of the incoherent Weber beam,
so that one can numerically find the full width of the half maximum (FWHM) of the main lobe of $|\gamma_{12}|^2$ to characterize the coherence property,
which is around several hundreds of nanometers and increases with the propagation distance.
From the solid curves in Fig. \ref{fig3}(e),
one can find that the beam width $x_0$ of the incoherent Weber beam is almost in the region $7~\mu{\rm m}<x_0<7.5~\mu{\rm m}$,
which is larger than the FWHM of $|\gamma_{12}|^2$.

In the end, we turn to the coherence length, as shown by the dashed curves in Fig. \ref{fig3}(e).
Corresponding to the main lobe of the incoherent beam,
the coherence length is really small and on the order of a single wavelength (hundreds of nanometers),
which is in accordance with the results presented in Figs. \ref{fig3}(b)-\ref{fig3}(d).
One trivial fact worth mentioning is that the incoherence will be strengthened if the transverse displacement extends over a larger region.

\section{Conclusion}\label{conclusion}

In conclusion, we have investigated the coherent and partially incoherent Weber beams.
For the coherent case, the nonparaxial accelerating Talbot effect cannot be obtained,
because the lobes of the Weber beam do not accelerate in unison.
While for the incoherent case the beam still maintains the nonparaxial accelerating property.
Both the coherence function and the coherence length are presented, to display the degree of coherence,
and we have found that the coherence length is on the order of a micrometer, which is comparable with the single wavelength.
The coherent and incoherent nonparaxial Mathieu accelerating beams are not discussed here,
because the research method reported in this paper is also applicable to such nonparaxial accelerating beams
and similar results are expected.
The only difference is that they accelerate along elliptical curves,
and are solutions of the Helmholtz equation in elliptical cylindrical coordinates.
This presentation enriches the understanding of nonparaxial accelerating beams and the coherence properties of such nonparaxial accelerating light.

\section*{Acknowledgements}
The National Basic Research Program of China (2012CB921804);
National Natural Science Foundation of China (61308015, 11474228);
Key Scientific and Technological Innovation Team of Shaanxi Province (2014KCT-10);
and Qatar National Research Fund  (NPRP 6-021-1-005).
MRB also acknowledges support by the Al Sraiya Holding Group.

\bibliographystyle{myprx}
\bibliography{my_refs_library}

\end{document}